\documentclass[sn-mathphys,Numbered]{sn-jnl}

\usepackage{graphicx}%
\usepackage{multirow}%
\usepackage{amsmath,amssymb,amsfonts}%
\usepackage{amsthm}%
\usepackage{mathrsfs}%
\usepackage[title]{appendix}%
\usepackage{xcolor}%
\usepackage{textcomp}%
\usepackage{manyfoot}%
\usepackage{booktabs}%
\usepackage{algorithm}%
\usepackage{algorithmicx}%
\usepackage{algpseudocode}%
\usepackage{listings}%
\usepackage{xfrac}

\def\ket#1{{\left\vert#1\right\rangle}}
\def\abs#1{{\left|#1\right|}}

\def\tFWHM{\tau_\text{FWHM}}
\def\tctrl{\tau^\text{ctrl}_\text{FWHM}}
\def\Deltau{\Delta\tau^\text{ctrl}}

\begin{document}

\title{High-efficiency, high-speed, and low-noise photonic quantum memory 
}


\author*[1,2]{\fnm{Kai} \sur{Shinbrough}}\email{kais@illinois.edu; vlorenz@illinois.edu}
\author[1,2]{\fnm{Tegan} \sur{Loveridge}}
\author[1,2]{\fnm{Benjamin D.} \sur{Hunt}}
\author[3]{\fnm{Sehyun} \sur{Park}}
\author[1,2]{\fnm{Kathleen} \sur{Oolman}}
\author[3]{\fnm{Thomas O.} \sur{Reboli}}
\author[3]{\fnm{J. Gary} \sur{Eden}}
\author*[1,2]{\fnm{Virginia O.} \sur{Lorenz}}

\affil[1]{\orgdiv{Department of Physics}, \orgname{University of Illinois Urbana-Champaign}, \orgaddress{\street{1110 W. Green St.}, \city{Urbana}, \state{IL} \postcode{61801}, \country{USA}}}

\affil[2]{\orgdiv{Illinois Quantum Information Science and Technology (IQUIST) Center}, \orgname{University of Illinois Urbana-Champaign}, \orgaddress{\street{1101 W. Springfield Ave.}, \city{Urbana}, \state{IL} \postcode{61801}, \country{USA}}}

\affil[3]{\orgdiv{Department of Electrical and Computer Engineering}, \orgname{University of Illinois Urbana-Champaign}, \orgaddress{\street{306 N. Wright St.}, \city{Urbana}, \state{IL} \postcode{61801}, \country{USA}}}


\abstract{We present a demonstration 
of simultaneous high-efficiency, high-speed, and low-noise operation of a photonic quantum memory. 
By leveraging controllable collisional dephasing in a neutral barium atomic vapor, we demonstrate a significant improvement in memory efficiency and bandwidth over existing techniques. 
We achieve greater than 95\% storage efficiency and 26\% total efficiency of 880 GHz bandwidth photons, with $\mathcal{O}(10^{-5})$ noise photons per retrieved pulse. These ultrabroad bandwidths 
enable rapid quantum information processing and contribute to the development of practical quantum memories with potential applications in quantum communication, computation, and networking.
}

\maketitle


Photonic quantum memories play a vital role in many quantum information processing applications by enabling the on-demand storage and retrieval of traveling qubits --- quantum states of light \cite{kimble2008quantum,hammerer2010quantum,sangard2011quantum,lvovsky2009optical,kok2007linear}. Achieving high-efficiency, high-speed, and low-noise operation of such memories is of utmost importance for the practical realization of these applications. In recent years, significant progress has been made in the development of efficient photonic quantum memories using various physical systems, including solid-state materials, cold atomic gases, and warm atomic vapors \cite{lvovsky2009optical,shinbrough2023broadband,simon2010quantum,cho2016highly}.

As we show, a trade-off exists in these atomic ensemble-based systems between memory bandwidth and storage efficiency, owing to the 
mismatch between broad photon bandwidths and typically narrow atomic linewidths. This trade-off presents a problem for high-speed photonic quantum information processing utilizing quantum memories, and for interfacing quantum memories with typically broadband quantum light sources based on parametric down-conversion or four-wave mixing, which are the workhorse for many quantum optics experiments \cite{zhong2020quantum,wang2016experimental,bouwmeester1997experimental,jennewein2000quantum,tittel1998violation,pan1998experimental}.  To overcome this limitation, we present a novel approach to atomic ensemble-based quantum memory that relies on homogeneous collisional broadening of an intermediate atomic state via noble gas perturbers at controllable pressure. This collisional broadening reduces linewidth-bandwidth mismatch and thereby enhances memory efficiency in the broadband regime.

The contributions of this work are twofold. First, we demonstrate a generic and scalable approach 
to enhancing memory efficiency in atomic-vapor quantum memories based on collisional broadening. This demonstration is performed in a new medium --- atomic barium vapor --- with an orbital $\Lambda$-system, and our approach leads to a measured storage efficiency of 95.6(3)\%. Second, we present a comprehensive characterization of our system's performance, including measurement of memory efficiency, lifetime, noise level, and full reconstruction of retrieved photon amplitude and phase. This characterization reveals a regime of operation for atomic ensemble memories we name Near-Off-Resonant Memory (NORM) operation, which leads to enhancements in total efficiency of around 5\% 
in our system. 
Throughout, we highlight the unique advantages of our choice of atomic species 
for this application, including the absence of four-wave mixing noise, the telecom-compatibility of the control field wavelength, and the small collisional dephasing rate and long natural lifetime of the chosen storage state (0.25 seconds in the bare atom \cite{migdalek1990multiconfiguration}). 
As a proof of principle, we store and retrieve weak coherent states with $\lesssim$1 average photon per pulse; expansion of this experiment to storage of single-photon Fock states is straightforward, and may be the subject of future work. The achievements of this work pave the way for the realization of practical, high-speed photonic quantum memories, with potential applications in quantum communication, computing, and networking.

In the following sections, we describe the principles behind our approach, the experimental setup, our measurement results, and we discuss the implications and future prospects of our work.

\section{Results}\label{Results_sec}

\subsection{Collisional broadening as a resource for improving broadband memory efficiency.}\label{coll_sec}

\begin{figure*}[h!]
	\centering
	\includegraphics[width=\linewidth]{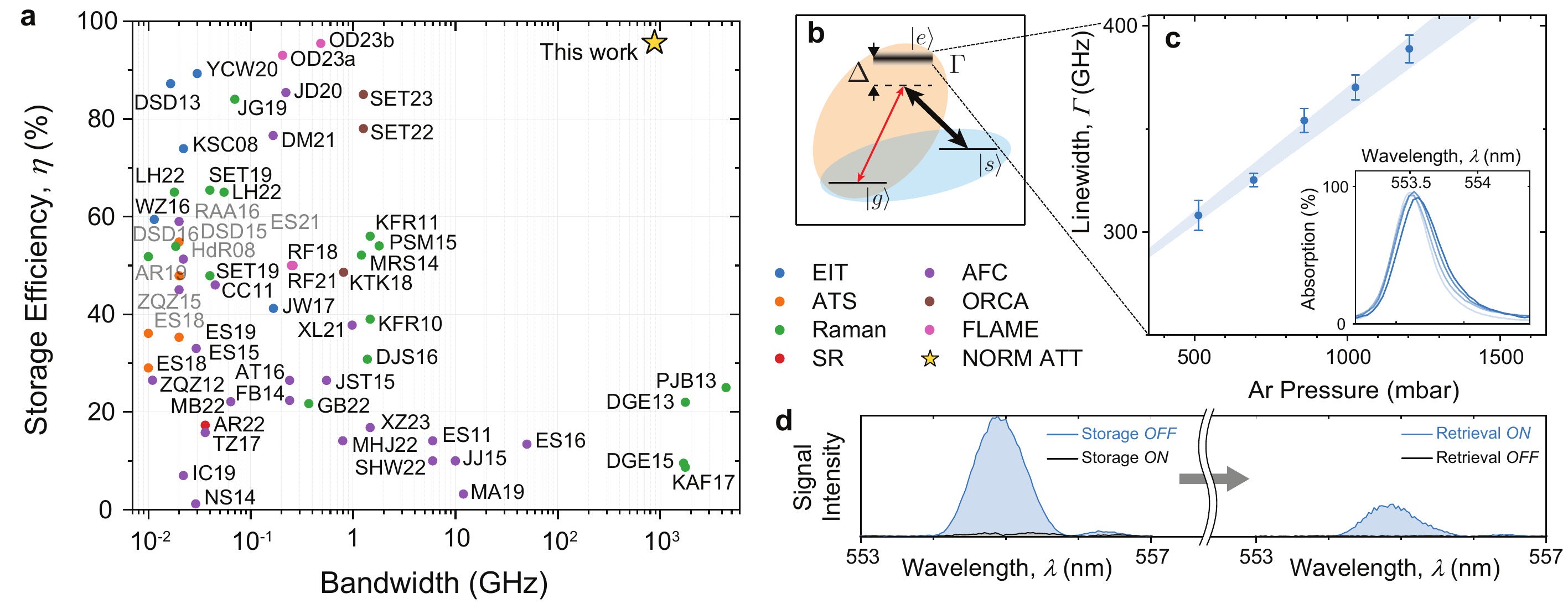}
	\caption{
 \textbf{Enhancement in memory efficiency due to collisional broadening.} \textbf{a}, Comparison of storage efficiencies and signal field bandwidths for atomic ensemble quantum memories in the broadband regime ($>10$ MHz). References are specified by first author initials and publication year; for complete reference information see Supplementary Information. 
    \textbf{b}, 
    Atomic energy level structure for $\Lambda$-type quantum memory ($\Delta$, detuning; $\Gamma$, excited state linewidth; red arrow, signal field; black arrow, control field). \textbf{c}, Measured collisionally broadened excited state linewidths for the $^1S_0$ $\rightarrow$ $^1P_1$ ($\ket{g}\rightarrow\ket{e}$) transition in barium as a function of argon (Ar) buffer gas pressure (inset: absorption spectra). \textbf{d}, Spectral waveforms of the signal field before storage (blue, left panel), after quantum 
    storage (grey, left panel), before retrieval (grey, right panel), and after retrieval (blue, right panel).}
	\label{fig1}
\end{figure*}

It is well-known that the resonant optical depth, 
$d$, 
of a three-level atomic-ensemble quantum memory 
sets an upper bound on the maximum achievable storage efficiency of the memory, of the form $\eta_\text{opt} \approx 1-2.9/d$ \cite{gorshkov2007universal,gorshkov2007photon_2,nunn2008multimode}. Optical depth, proportionate to atom number, is therefore a resource for increasing memory efficiency. Here we propose and demonstrate experimentally another resource for atomic-ensemble quantum memories: the intermediate-state homogeneous linewidth. The value of this resource is most clear when considering the so-called absorb-then-transfer (ATT) memory protocol \cite{moiseev2001complete,vivoli2013high,gorshkov2007photon_2,carvalho2020enhanced,shinbrough2023broadband,shinbrough2021optimization}. In this quantum storage protocol, resonant linear absorption maps a photonic qubit---the signal field---onto an atomic polarization of the form 
$P\sim\sum_{j=1}^{N} b_je^{ik_s z_j}\ket{g_1\cdots e_i\cdots g_N},$
where the sum runs over atoms 1 to $N$, each with a spatially dependent amplitude ($b_j$) and phase ($k_s z_j$). This collective Dicke state involving the ground state $\ket{g}$ and the excited state $\ket{e}$ is then mapped via a $\pi$-pulse optical control field onto a long-lived ``spin-wave'' state of the form 
$B\sim\sum_{j=1}^{N} c_je^{i(k_s - k_c)z_j}\ket{g_1\cdots s_i\cdots g_N},$ 
involving the storage state $\ket{s}$ (see Fig.~\ref{fig1}b). The protocol thus has two stages: resonant linear absorption, and $\pi$-pulse population transfer. The efficiency of the second stage is ensured simply by accurate tuning of the control field pulse area, but the efficiency of the first stage depends critically on both the resonant optical depth and linewidth of the $\ket{g}\rightarrow\ket{e}$ 
transition. A complete discussion of the dependence of linear absorption on these two memory parameters can be found in Ref.~\cite{shinbrough2023broadband}, but the fundamental physical intuition is clear: When the signal field bandwidth is broader than the transition linewidth, the frequency components of the signal field outside of the linewidth are not absorbed by the ensemble, and therefore contribute to transmission loss or memory inefficiency. This loss can be compensated 
by either increasing optical depth or increasing the transition linewidth, but in the ultra-broadband regime where the signal bandwidth is much greater than than the atomic linewidth ($\delta_s\gg\gamma$), increasing linewidth is significantly more 
effective than increasing optical depth. 
A larger homogeneous linewidth is also a resource for the resonant protocols of electromagnetically induced transparency (EIT) and Autler-Townes Splitting (ATS), but whether a given memory is `linewidth limited' or `optical-depth limited' in general depends on the specific parameters of the system.

Another factor motivates the use of homogeneous linewidth broadening as a resource for improving memory efficiency: Optical transitions in warm atomic vapors are typically dominated by Doppler broadening, which introduces inhomogeneous dephasing during the storage and retrieval operations and therefore decreases the coherent reemission probability of the memory. 
Introducing an intentional source of homogeneous broadening larger than the Doppler linewidth removes this source of 
memory inefficiency while simultaneously increasing memory bandwidth and decreasing protocol time. In general, this approach may also reduce memory lifetime, as the storage state may also undergo collisional broadening beyond its Doppler linewidth. In our experiment however, the collisional cross-section of the atoms in the storage state is at least an order of magnitude smaller than the atoms in the excited state. Therefore, for a fixed buffer gas pressure the collisional broadening of the excited state is significantly larger than the collisional broadening of the storage state. Experimentally, we are able to demonstrate Doppler-limited memory lifetimes while simultaneously taking advantage of large collisionally broadened excited state linewidths (see Sec.~\ref{char_sec} and Methods). 

In order to harness linewidth-broadening as a resource, we begin with a Doppler-broadened vapor of neutral barium in a home-built heat pipe oven vapor cell (described further in Methods). We introduce argon buffer gas perturbers into the vapor cell at controllable pressure, with a ratio of roughly $10^4$ argon atoms per gaseous barium atom. The argon perturbers interact with the barium atoms via impact broadening at timescales longer than $\mathcal{O}(1-10)$ ns, depending on argon pressure, and interact via quasi-static broadening at shorter timescales (further details in Methods). In both cases, 
the collisional broadening of the intermediate $^1P_1$ excited state in the barium atoms is homogeneous and of order $\Gamma = 100$ GHz (full width at half maximum), significantly in excess of the 1-10 GHz temperature-dependent Doppler linewidth. We perform a modified version of the ATT 
protocol outlined above, where both signal and control fields are detuned $\Delta = 5\Gamma$ below resonance. This modification ensures less than 1\% absorption of the signal field in the absence of the control field, 
such that when the control field is turned on, any increase in absorption (or decrease in transmission) is attributable to quantum storage. 

We summarize the key results of this demonstration in Fig.~\ref{fig1}. Fig.~\ref{fig1}a presents a direct comparison of the storage efficiency and bandwidth achieved with our approach with the storage efficiencies and bandwidths achieved with existing techniques based on either lifetime-broadened, Doppler-broadened, or inhomogeneously broadened three-level quantum memories. 
These existing techniques employ a variety of different storage protocols [colors in Fig.~\ref{fig1}a], but in all cases the region of high-efficiency operation has to date been limited to $\le\mathcal{O}(1)$ GHz. In the ultrabroadband regime investigated in this work, storage efficiencies have previously been limited to roughly 25\%. We attribute the significant increase in storage efficiency demonstrated in this work to the use of a collisionally broadened linewidth [Fig.~\ref{fig1}b-c] as an additional resource. Fig.~\ref{fig1}c shows measured collisionally broadened linewidths in our system between 300 and 400 GHz, centered on the 553.5 nm $^1S_0$ $\rightarrow$ $^1P_1$ ($\ket{g}\rightarrow\ket{e}$) transition in barium, that are linearly dependent upon argon (Ar) buffer gas pressure as expected for collisional broadening. Fig.~\ref{fig1}d shows the raw spectrally resolved storage and retrieval data in our system, from which we extract $95.6\pm0.3$\% storage efficiency and $26\pm1$\% total (end-to-end) efficiency near zero time delay. 
The total efficiency of our memory is limited by available control field power and can be improved significantly with a higher pulse energy control field. We note that a trend similar to the one shown in Fig.~\ref{fig1}a also exists for total efficiency as a function of signal bandwidth, though the trend 
is 
considerably noisier, in part due to system-specific inefficiencies such as poor phasematching and insufficient available control field power. 

\subsection{Memory characterization and performance.}\label{char_sec}

\begin{figure}[h!]
	\centering
	\includegraphics[width=0.5\columnwidth]{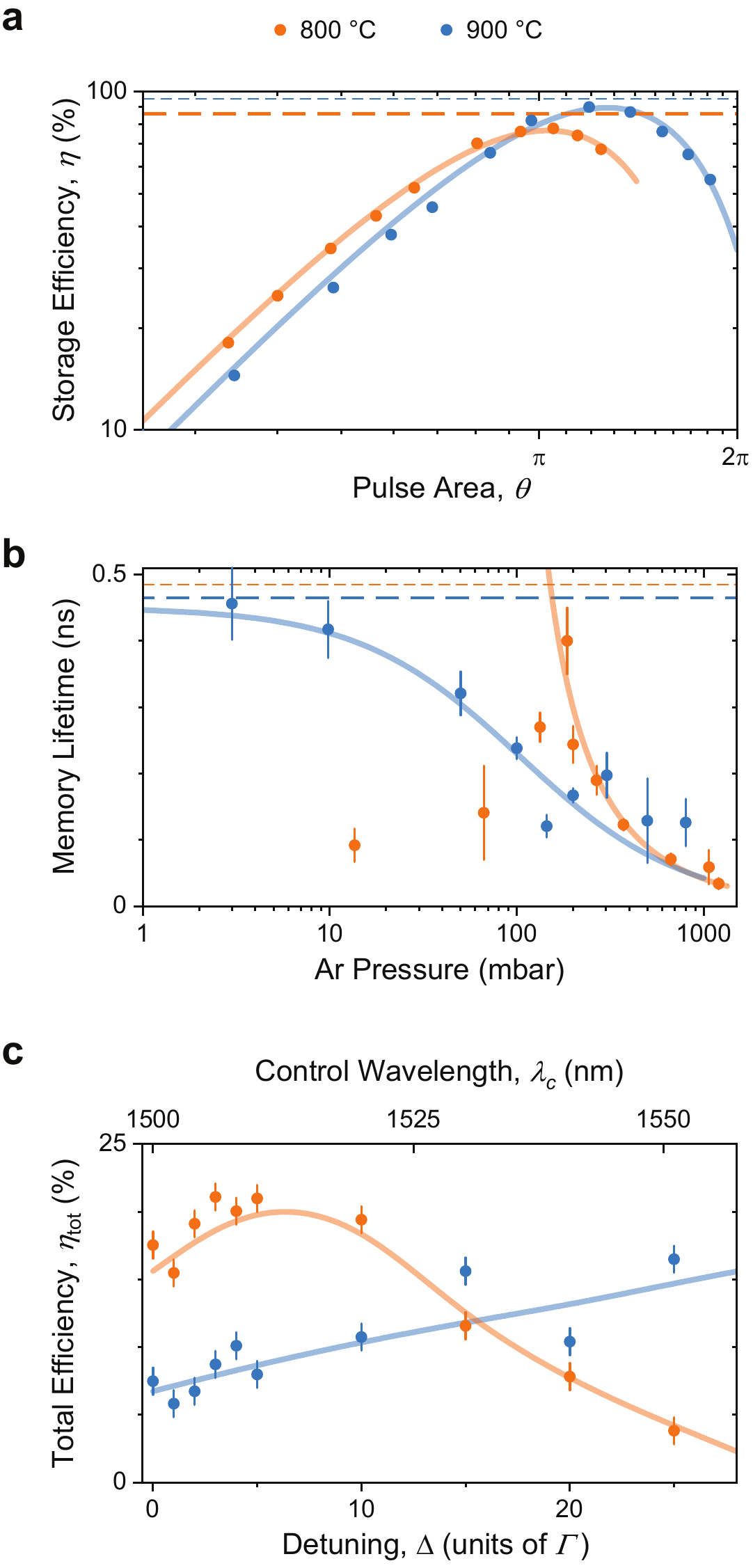}
	\caption{\textbf{Telecom-compatible 
    memory characterization.} 
    \textbf{a}, Optimization of storage efficiency with respect to control field pulse area at 800 $^\circ$C (orange) and 900 $^\circ$C (blue). Horizontal lines represent the 
    theoretical optimal bound on storage efficiency at each temperature. Statistical errors are smaller than the marker size. Solid curves are numerical fits based on the Maxwell-Bloch equations (see Methods). \textbf{b}, Memory lifetime as a function of argon (Ar) buffer gas pressure. Horizontal lines represent the limit on memory lifetime set by Doppler or motional dephasing 
    at each temperature. Error bars represent statistical error propagated through a decay model fit (see Methods). Solid curves are fits to an inverse function representing collisional broadening. \textbf{c}, 
    Total (end-to-end) memory efficiency measured as a function of detuning, showing near-off-resonant memory (NORM) operation. 
    Error bars represent systematic error due to control field power variation.  Solid curves: numerical fit to Maxwell-Bloch model (see Methods).
 }
	\label{fig2}
\end{figure}

In addition to the near-resonant storage and retrieval demonstrated in Fig.~\ref{fig1}d, we perform several experiments to characterize the performance of our photonic quantum memory. First, we measure the response of the memory to increasing control field power during the storage operation. Shown in Fig.~\ref{fig2}a, at 800 $^\circ$C we observe a maximum in storage efficiency near $\pi$ control field pulse area, as expected for the ATT protocol. At 900 $^\circ$C, with higher optical depth and larger collisional broadening we are able to achieve larger control field pulse areas with the same total available control field power, and we observe an optimal control field pulse area of $1.25\pi$. 
The experimental data in Fig.~\ref{fig2}a are fit to a numerical model based on the Maxwell-Bloch equations (see Methods) for Ar pressures of 670 and 13 mbar for 800 and 900 $^\circ$C, respectively. The 
dashed 
horizontal lines 
represent the optimal bound on storage efficiency for a given temperature and independently measured 
optical depth ($\eta_\text{opt}$). We achieve near saturation of this bound after a half Rabi oscillation, confirming the coherence 
of our memory and the applicability of the Maxwell-Bloch model.

Next, we turn to the coherence lifetime of our memory. 
The collisional broadening we employ is not intrinsically state-selective; in addition to collisional broadening of the intermediate excited state, the presence of noble gas perturbers also leads to collisional broadening of the metastable or storage state. The linewidth of this state determines the coherence time of our memory, and as such a tradeoff exists between increasing memory efficiency and maintaining a long coherence time. To this end we measure the coherence lifetime of our memory as a function of argon buffer gas pressure, shown in Fig.~\ref{fig2}b. The horizontal lines in Fig.~\ref{fig2}b represent the limit to memory lifetime imposed by Doppler broadening at each temperature, due to the thermal motion of barium atoms which decoheres the spatially varying phase of the spin-wave. We observe qualitatively different behavior for the two vapor cell temperatures investigated in this work. 
At 800 $^\circ$C, we observe a short memory lifetime at low argon pressure, which then increases to a maximum around 200 mbar before decaying according to an inverse model indicative of collisional broadening. At low pressures, we believe the memory lifetime may be reduced due to the formation of a sub-ensemble of weakly bound barium-argon molecules \cite{czuchaj1998calculation,bezrukov2019ab,buchachenko2018interaction}, but further work is needed to investigate this effect. At 900 $^\circ$C, we observe a memory lifetime that asymptotes to the Doppler limit at low pressure and that follows the expected collisional model for increasing argon pressure. As the memory efficiency is unchanged for this range of argon pressures, we 
are able to achieve Doppler-limited memory lifetimes while still benefiting from the enhancement to memory efficiency due to collisional broadening. The reason for this is an asymmetry in the collisional cross sections of the $^1P_1$ excited state and the $^1D_2$ storage state, where the excited state collisional cross section is significantly larger. This means that for fixed argon pressure, the excited state will experience significantly larger collisional broadening than the storage state, thus allowing for high-efficiency memory operation without a significant trade off in memory lifetime. The memory lifetimes demonstrated in this work are almost a factor of 2 longer than previous ultrabroadband photonic quantum memories \cite{england2013photons,england2015storage,fisher2017storage,bustard2013toward}. We note that the radiative lifetime of the $^1D_2$ state in the bare atom is 0.25 sec \cite{migdalek1990multiconfiguration}, which leaves significant room for improvement of our memory lifetime using recently developed dephasing-protection techniques \cite{finkelstein2021continuous}, whereas previous ultrabroadband memories have been limited by more fundamental constraints on memory lifetime \cite{england2013photons,england2015storage,fisher2017storage,bustard2013toward}.

We also measure the carrier-frequency dependence of our memory. Keeping the signal and control fields in two-photon resonance, we vary the detuning, $\Delta$, shown in Fig.~\ref{fig1}b, and measure the total, end-to-end efficiency of our memory. At both temperatures we observe a maximum total efficiency at non-zero detuning, an effect we name Near-Off-Resonant Memory (NORM) operation. This regime of memory operation occurs when the adiabaticity of the control field is 
less than the adiabaticity of the memory set by the optical depth, excited state linewidth, and signal bandwidth (for more details, see Methods). Most often, as in our experiment, this occurs when control field power is limited. We note this adiabaticity criterion is a sufficient, but not necessary condition for NORM operation as many sets of control field parameters do not possess a well-defined adiabaticity according to our definition. 
In this regime it is beneficial to reduce the light-matter coupling of the signal field and atomic ensemble by detuning the signal field slightly off resonance. This operation regime has been observed experimentally in previous work \cite{finkelstein2018fast,wolters2017simple}, but we explore the effect theoretically for the first time, and offer an intuitive explanation in the Methods section. Our model fit is in good agreement with the data in Fig.~\ref{fig2}c, where we observe an optimal detuning around $\Delta=5\Gamma$ at 800 $^\circ$C (190 mbar) and 
$\Delta>25\Gamma$ at 900 $^\circ$C (270 mbar). At 900 $^\circ$C we note the enhanced memory efficiency at 1550 nm control field wavelength, which highlights the telecom-compatibility of this memory. With optical pumping of the initial population into the storage state, we can in principle implement the same efficiency storage and retrieval at 1550 nm signal field wavelength. At 1550 nm, blackbody radiation of the heat pipe oven will introduce additional noise, but experimentally we measure blackbody radiation below the current noise level, 
which is dominated by two-photon absorption of the control field and fluorescence (see Sec.~\ref{noise_sec}). Theoretically blackbody radiation noise can be reduced to arbitrarily low levels by increasing the distance between heat pipe and collection optics (thereby decreasing the collected solid angle). 

\subsection{Single-photon-level retrieval with full amplitude and phase reconstruction.}\label{noise_sec}

\begin{figure}[t]
	\centering
	\includegraphics[width=0.5\columnwidth]{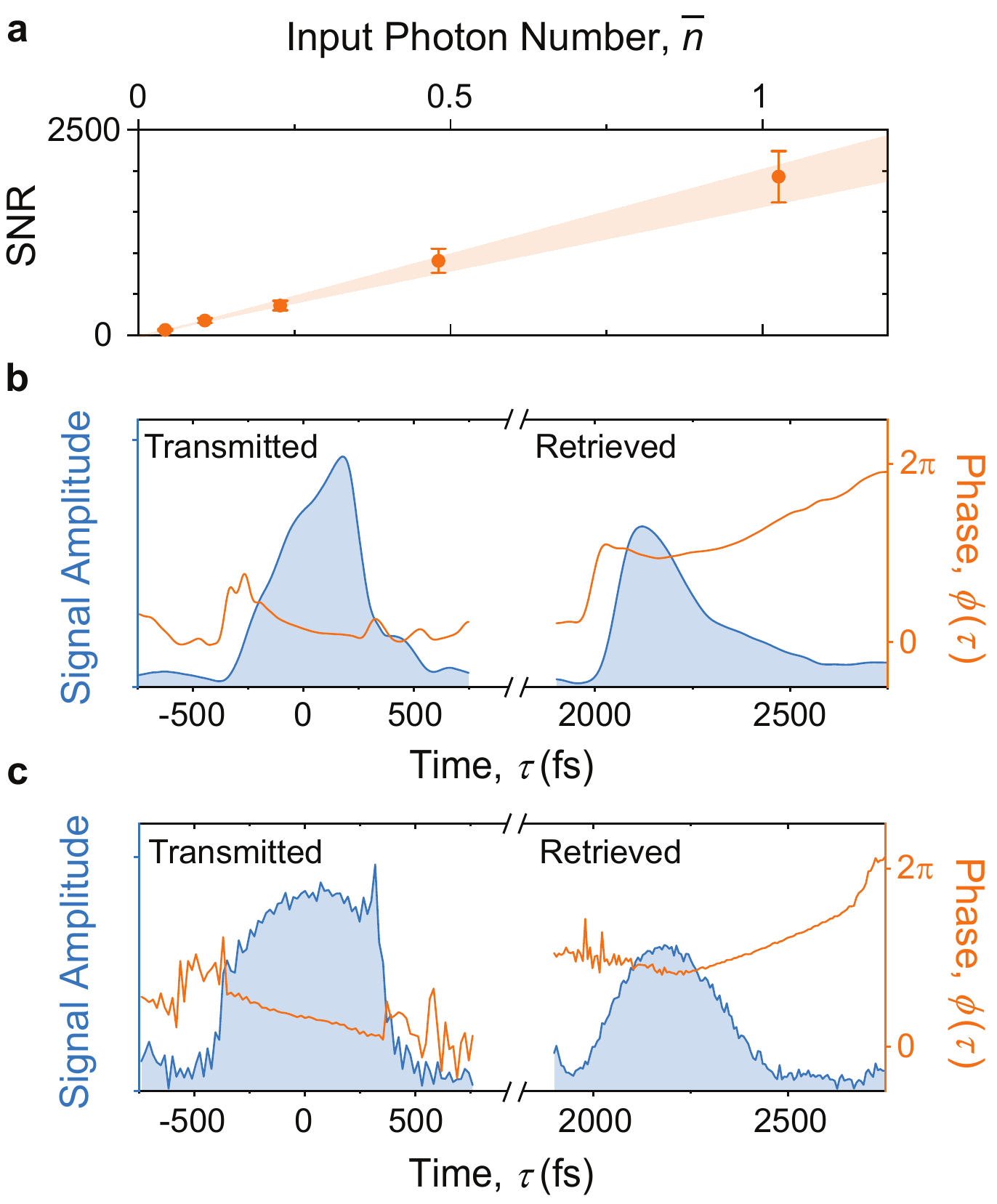}
	\caption{
    \textbf{Noise performance and full signal field amplitude and phase reconstruction.} \textbf{a}, 
    Measured ratio of average retrieved photon number to average noise photon number (signal-to-noise ratio, SNR) \textbf{b}, Predicted amplitude and phase of photonic field after the memory, including transmitted (left) and retrieved (right) components. \textbf{c}, Experimental amplitude and phase of the photonic field after the memory, reconstructed from spectral interference with a known reference.
 }
	\label{fig3}
\end{figure}

We now turn to the noise performance of our memory. Most $\Lambda$-type quantum memories are limited in noise performance by four-wave mixing (FWM) noise, wherein the control field coupling the storage and excited states also acts off-resonantly along the ground to excited state transition, generating spurious Stokes and anti-Stokes photons. The anti-Stokes photons generated in this process overlap exactly with the retrieved signal field in time, frequency, polarization, and spatial mode, but carry none of the quantum information stored in the original signal field. Several techniques have been developed to mitigate FWM noise in $\Lambda$-type quantum memories \cite{thomas2019raman,bustard2016reducing,nunn2017theory,saglamyurek2019single}, typically at the expense of additional optical fields, cavities, or more complex beam routing. By contrast, our barium $\Lambda$-type quantum memory is intrinsically FWM noise free. This is due to the large, 340 THz ground-storage state splitting of the $^1S_0$ and $^1D_2$ states. As this splitting is larger than the excited-storage state splitting ($^1P_1$-$^1D_2$, 200 THz) defining the resonant control field frequency, to first order the control field does not possess sufficient energy per photon to excite FWM noise. In Fig.~\ref{fig3}a we show the measured signal-to-noise ratio of our memory -- defined as the ratio of the average retrieved signal field photon number to the average noise photon number --- as a function of the average input photon number of a weak coherent state. 
We fit this measured data to a linear function in order to extract a signal-to-noise ratio of $\text{SNR}=1800$ at an average of 1 input photon per pulse. This represents a retrieved single-photon fidelity of $\mathcal{F} = 1-1/(\text{SNR}+1) = 0.9994$, which is the highest noise-limited fidelity of any $\Lambda$-type quantum memory to date \cite{shinbrough2023broadband}. We believe the noise performance of our memory is limited by two-photon absorption of our control field and fluorescence from either a high-lying atomic orbital or the silica windows of our heat pipe oven, both of which are near the dark count rate of our detectors.

The ultra-low noise performance of our memory 
combined with its ultrabroad bandwidth allows us to perform a novel fidelity characterization experiment. In Figs.~\ref{fig3}b and \ref{fig3}c we show the predicted and reconstructed amplitude and phase of our output signal field, respectively. The predicted amplitude and phase is generated via integration of the Maxwell-Bloch equations with an input signal field generated via the Fourier transform of our measured signal spectrum, assuming a flat spectral phase (see Methods).  The output field has two components, the field transmitted after the storage control field pulse, and the field retrieved via the retrieval pulse. The amplitude and phase in Fig.~\ref{fig3}c are reconstructed from measured spectral interferograms between the output field and a known reference (see Methods). This spectral interference technique relies on the low-noise and broad bandwidth operation of our memory that make high-resolution measurement of spectral interference experimentally feasible. The agreement between the output field amplitude and phase in Figs.~\ref{fig3}b and \ref{fig3}c is imperfect due to the limited stability and acquisition time of our interferometer, the assumption of a flat input phase, and the resolution of our spectrometer, but the agreement is sufficient to accurately extract linear and quadratic components of the retrieved signal field temporal phase, which are necessary for spectral-temporal compression and shaping of the retrieved field, which may be necessary for some applications.

\section{Discussion}\label{Discussion_sec}

In this work, we have demonstrated a novel and scalable approach to enhancing atomic-vapor quantum memory efficiency via tunable collisional broadening. This approach overcomes the trade-off between memory bandwidth and storage efficiency present in the literature, constituting a significant advance in photonic quantum memory performance. Our barium-based quantum memory exhibits simultaneously near optimal storage efficiency, a Doppler-limited lifetime, a telecom-wavelength control field, and the highest noise-limited fidelity of any $\Lambda$-type quantum memory to date. Our memory operates in a Near-Off-Resonant Memory or NORM regime, which balances resonant reabsorption loss and weak off-resonant light-matter coupling, and the absence of four-wave mixing noise combined with the broad bandwidth of our signal field 
allows for full amplitude and phase reconstruction of the output signal field via spectral interference. Taken as a whole, this performance makes our memory a promising candidate for applications in quantum communication, multiphoton state preparation, and local quantum processing, but further work is needed to transform our memory into a practically useable device. We discuss these limitations and future improvements here.

The most important bottleneck for broadband ensemble-based quantum memories lies in their limited storage lifetime \cite{shinbrough2023broadband}. A common figure of merit often quoted for ensemble-based memories is the time-bandwidth product --- $\text{TBP} = T\times BW$, where $T$ is the $1/e$ memory lifetime and $BW$ is the signal field bandwidth. TBP represents the memory lifetime in multiples of the pulse duration, and is therefore a rough estimate of how many operations could be performed in a photonic quantum processor during the storage time at a clock rate comparable to the signal bandwidth. The time-bandwidth product in this work of $\text{TBP} = 980$ is comparable to the state-of-the-art for broadband ensemble memories \cite{reim2011single,wei2020broadband,heller2022raman,bustard2013toward}, but this simple calculation conceals the fact that clock rates beyond a few GHz are not compatible with contemporary fast electronics and electro-optics, and may not be usable in a practical device. The time-clock-rate product $\text{TRP} = T\times R$, where $R$ is clock rate, represents the number of clock cycles for which a quantum memory can store a photonic qubit. For our memory the TRP is only $\sim1$ for a standard 2 GHz CPU clock, and is even lower for most quantum photon pair source repetition rates, which are typically in the MHz range \cite{lounis2005single,eisaman2011invited,couteau2023applications}. 
This means that in a quantum processor with 2 GHz clock rate, our memory can only store a photonic qubit for 1 clock cycle. A TRP equal to (but ideally much greater than) 1 is a prerequisite for a useful memory device, and our 
memory is the first in the ultrabroadband regime to meet this threshold. As noted in Sec.~\ref{char_sec}, our barium memory has significant room for improvement in memory lifetime ($T$), and therefore in TRP. The lifetime of barium in the bare atom is 0.25 sec \cite{migdalek1990multiconfiguration}; if our memory lifetime could saturate this bound, this would correspond to a 2-GHz TRP of $5\times10^8$. Recent work on mitigating Doppler dephasing in atomic ensemble quantum memories \cite{finkelstein2021continuous} has direct application to this work, and may lead to significant improvement of our TRP. Our memory TRP could also be improved by increasing the clock-rate $R$; currently, broadband single-photon sources tend to operate at 1-10 MHz count rates, often limited by detector saturation, but in principle future improvements could bring these count rates up to the $\mathcal{O}(100)$ GHz bandwidth of our memory. With these two improvements (longer memory lifetime and higher repetition rate sources), the TRP for our memory could in principle improve to a bound of $\mathcal{O}(10^{10})$. Instead of using the $1/e$ lifetime in calculating TRP, which is a defined relative to the end-to-end efficiency at zero storage time, one could further consider an absolute lifetime in the calculation of TRP, e.g., the lifetime defined by 90\% absolute end-to-end efficiency, which is a considerably more demanding metric (no broadband quantum memory to-date can achieve $>$1 TRP with this threshold).


In the context of scalability, one practical limitation of this work 
in particular is the use of a high-temperature heat pipe oven vapor cell with large spatial extent ($\sim$1 ft) and power consumption (1000 W) (see Methods). Cold atomic ensemble quantum memories tend to have similar 
practical limitations, but miniaturization of these experiments is in principle possible \cite{du2004atom,folman2000controlling,elliott2018nasa,becker2018space}. 
Due to the low vapor pressure and high melting temperature of barium, the generation of a thermal vapor of sufficient optical depth requires long propagation lengths and/or temperatures above 900 C, for which creating long-lifetime vapor cells is difficult \cite{lorenz2008high}. Light-induced or electrically-induced atomic desorption \cite{alexandrov2002light,karaulanov2009controlling} and laser ablation \cite{matsuo1999radiative,harilal2003internal} are alternative methods that may be used to generate a dense cloud of barium vapor, and may be more amenable to miniaturization, though the optical depths generated in these processes tend to be low ($<$10). Work on a compact, high-density, low-power source of atomic barium is on-going.

In conclusion, we have experimentally demonstrated a simultaneously high-efficiency, high-speed, and low-noise atomic ensemble quantum memory. Our approach to increasing memory efficiency via tunable collisional broadening is resource-efficient, and leads to record efficiencies in the ultrabroadband regime. 
We have carefully considered the strengths and limitations of this work, and have suggested avenues for future improvement. With these improvements, the quantum memory developed in this work may serve as a critical enabling technology for quantum applications in communication, metrology, and computing.


\backmatter

\bmhead{Acknowledgments}

This work was supported by NSF grant Nos. 1640968, 1806572, and 2207822; and NSF Award DMR1747426. We thank Andrey Mironov and Kavita Desai for helpful discussion related to barium-argon molecule formation; Ran Finkelstein, Eilon Poem, and Ofer Firstenberg for helpful discussion related to mitigation of Doppler dephasing; Yujie Zhang and Dong Beom 
Kim for helpful discussion related to the apparatus and measurement; and Ernest Northern and Jim Brownfield for expert machining of the heat-pipe oven.

\section{Methods}\label{Methods_Sec}

\subsection{Open-ended barium heat pipe oven.}

We employ a home-built stainless steel open-ended heat pipe oven \cite{vidal1969heat,vidal1971heat,vidal1996vapor} 
to generate a neutral ensemble of atomic barium in the presence of controllable argon buffer gas pressure (0-1333 mbar). The heat pipe is loaded with natural abundance solid barium metal fragments (American Elements) at room temperature under argon atmosphere, which melt and vaporize when the heat pipe is brought to 800-900 $^\circ$C (Ba melting point: 727 $^\circ$C) via an external resistive heater. A stainless steel mesh wick is inserted into the main chamber of the heat pipe to ensure convective flow and to prevent ``hot spots.'' The two regions of the chamber before the windows are water cooled to 18 $^\circ$C, and two \sfrac{1}{4} inch diameter apertures are inserted into the chamber, one on each end, to reduce the flow of barium vapor toward the windows. The heated region of the oven is 12 inches in length, and is surrounded by three clamshell resistive heaters. Two heaters with power consumption of 218 W heat the top of the heat pipe, and one heater with 600 W power consumption heats the bottom of the heat pipe.

\subsection{Experimental setup.}\label{exp_setup_sec}

\begin{figure}[t]
	\centering
	\includegraphics[width=0.4\columnwidth]{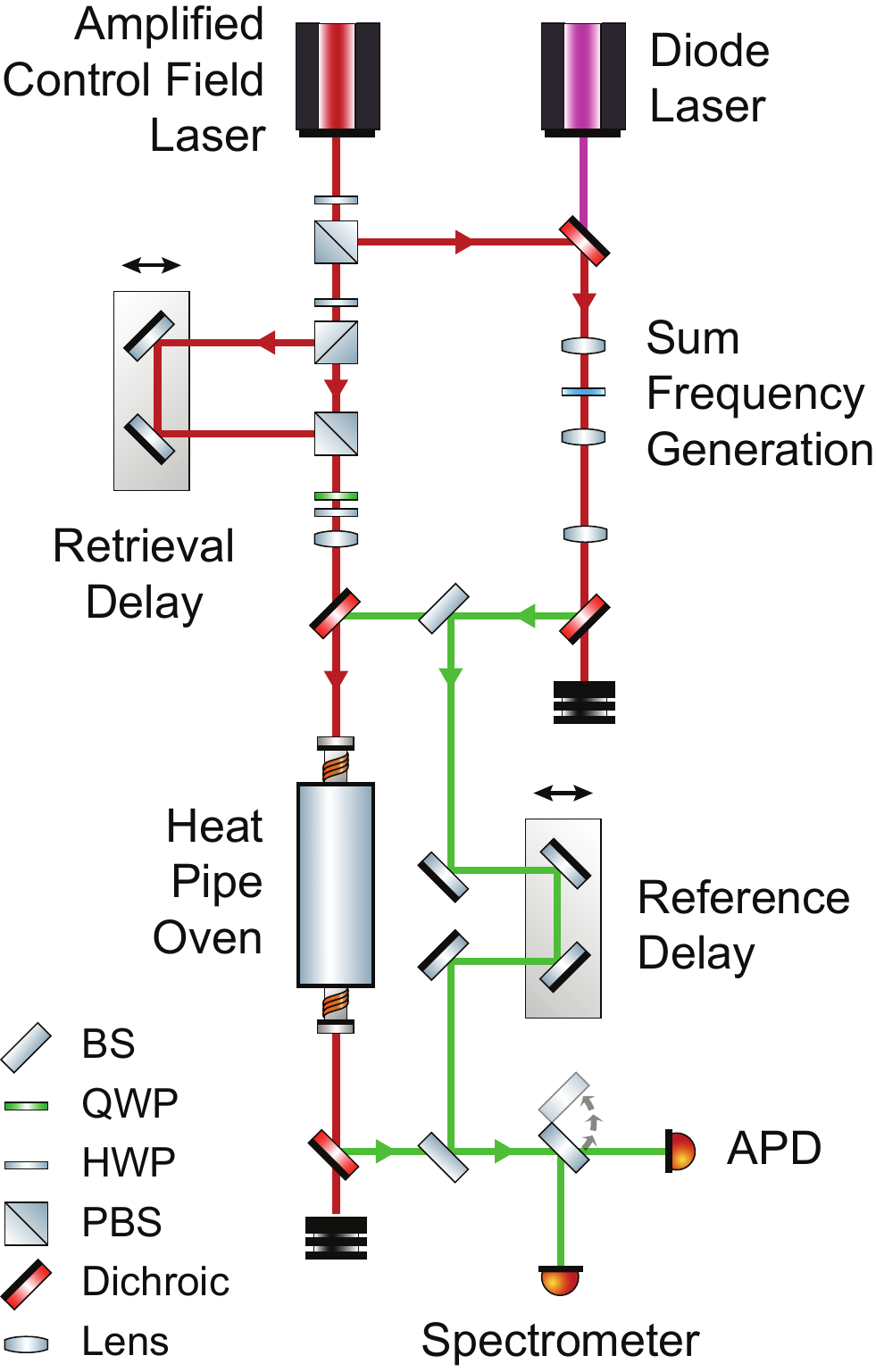}
	\caption{\textbf{Experimental schematic.}
    Simplified diagram of the barium quantum memory experiment. BS: Beam Splitter; QWP: Quarter Wave Plate; HWP: Half Wave Plate; PBS: Polarizing Beam Splitter; APD: Avalanche Photodiode.
 }
	\label{fig4}
\end{figure}

We provide a simplified experimental diagram in Fig.~\ref{fig4}. We employ an $\mathcal{O}(1)$ mJ pulse energy, $\mathcal{O}(100)$ fs, 1 kHz repetition rate, 800 nm Ti:Sapphire amplified laser system (Spectra-Physics) cascaded with a tunable white-light seeded amplified wavelength converter (Light Conversion) to produce $\mathcal{O}(100)$ uJ control field pulses between 1400 and 1700 center wavelength. We use a frequency-resolved optical gating (FROG) device (Mesa Photonics) to verify our control field pulses are Fourier-transform limited. A small fraction of this control field is split off and used to generate our signal field via sum-frequency generation with an 877 nm continuous wave diode laser in a room-temperature $\beta$-barium-borate (BBO) crystal (Newlight Photonics). The phasematching function of the sum-frequency generation process sets the signal field spectrum and 880 GHz full-width at half-maximum (FWHM) bandwidth (500 fs Fourier-limited FWHM duration). The control field is split into two pulses with controllable delay (retrieval delay in Fig.~\ref{fig4}) before being focused and overlapped with the signal field on a dichroic mirror (Semrock). The signal field is also split into two pulses, one of which is sent to the heat pipe while one is reserved to act as a reference for the spectral interference measurements. The signal and control field waist radii in the center of the heat pipe oven are 109(3) and 247(4) $\mu$m, respectively. After the heat pipe, the signal field is split from the control field with a dichroic mirror and 4 cascaded interference filters (Semrock), each with $>$93\% transmission at the signal field wavelength, before being recombined with the reference field and coupled into single-mode fiber. The in-fiber signal field is sent to either a high-quantum-efficiency spectrometer (Oxford Instruments, Andor) or a Si avalanche photodiode (Excelitas) and time-to-digital converter (ID Quantique).


\subsection{Collisionally broadened $^1P_1$ state linewidth measurements.}

We perform three separate spectroscopic measurements of the $^1S_0$ to $^1P_1$ transition at varying temperature and argon buffer gas pressure: white-light spectroscopy, scanning narrowband spectroscopy (147 MHz bandwidth, 3 ns probe), and coherent femtosecond spectroscopy (4.4 THz, 100 fs probe). The optical depth and collisionally broadened linewidth extracted from each method agree within measurement error with each other and with previous work \cite{ehrlacher1993noble}. Measured peak optical depths range from 25 to 50, depending on the heat pipe temperature, and measured $^1P_1$ state linewidth is $\mathcal{O}(100)$ GHz, and depends linearly on argon buffer gas pressure, as shown in Fig.~\ref{fig1}c. The equivalent optical depth of our system in a natural-linewidth ($\Gamma_\text{nat} = 120$ MHz \cite{dzuba2006calculations}) atomic ensemble (the so-called `cold OD') is roughly $d\Gamma/\Gamma_\text{nat} = 10^5$.

From kinetic theory \cite{crank1979mathematics,hirschfelder1964molecular} and the Van der Waals radii of barium and argon, we calculate the barium-barium and barium-argon diffusion coefficients, mean free path, and mean time between collisions for varying argon pressure and system temperature. For the experimental settings in this work (argon pressures between 1-1000 mbar and temperatures 800-900 $^\circ$C), the expected mean time between collisions is $\mathcal{O}(1-10)$ ns. At longer timescales probed via white light spectroscopy and scanning narrowband spectroscopy, we probe the impact broadening regime where phase discontinuities in the time domain emission (or absorption) of the radiating dipoles accounts for the broadened spectral line, whereas for the shorter timescales probed via coherent femtosecond spectroscopy, we probe the quasistatic broadening regime where the presence of many perturbers requires an averaging over the atom-perturber potential energy surface, and therefore line broadening \cite{corney1978atomic,czuchaj1998calculation,bezrukov2019ab,buchachenko2018interaction,allouche1995theoretical,lebeault1998vibrational}.

\subsection{$^1D_2$ coherence lifetime measurements.}\label{lifetime_sec}

\begin{figure}[t]
	\centering
	\includegraphics[width=0.5\columnwidth]{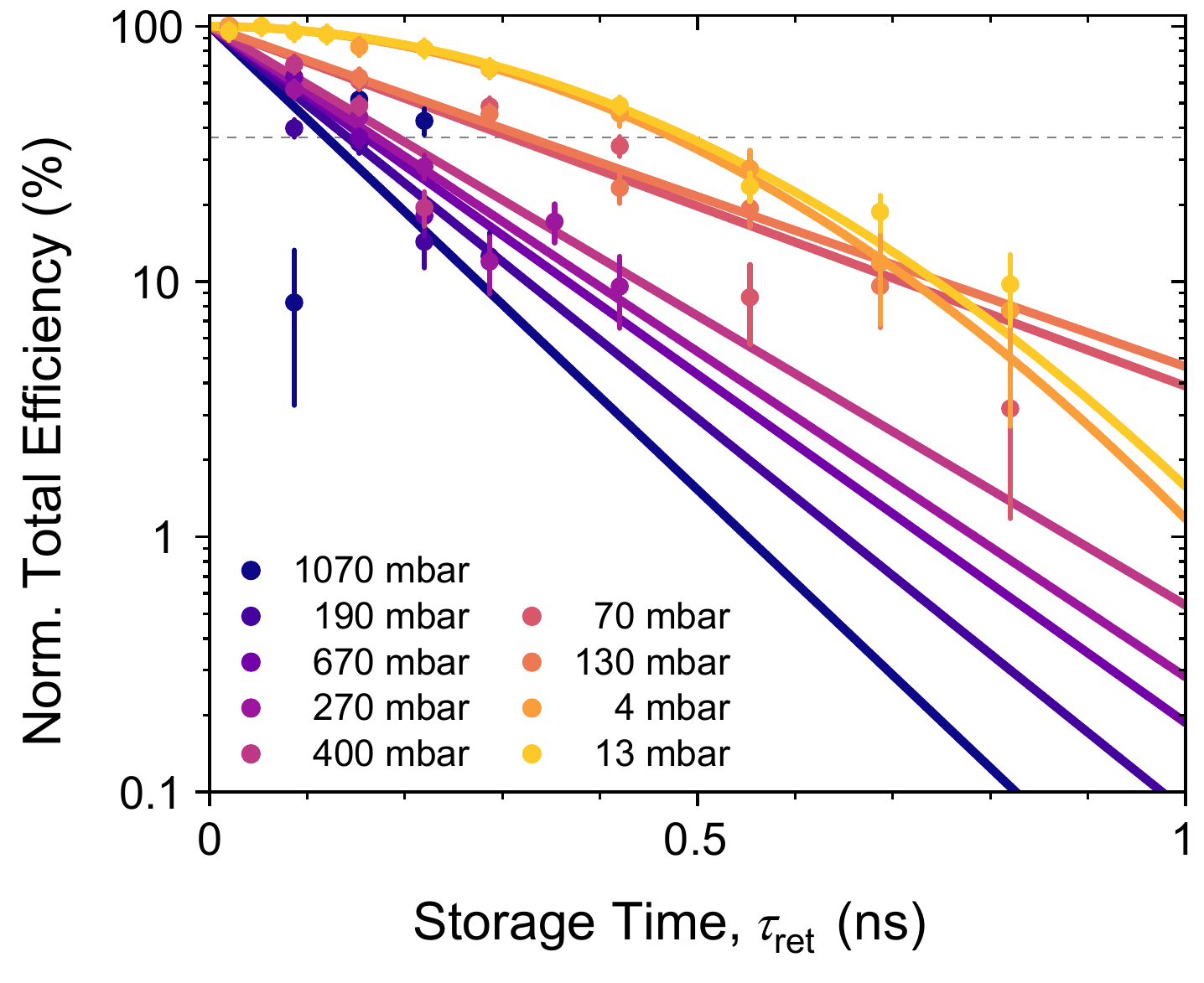}
	\caption{
 \textbf{Coherence lifetime measurements.} 
 Coherence lifetime measurements for the barium $^1S_0$--$^1D_2$ ensemble superposition state at 900 $^\circ$C for the varying argon buffer gas pressures in Fig.~\ref{fig2}b. Measured efficiencies are normalized to a maximum value of 100\%. Storage time refers to the time delay between storage and retrieval control field pulses, set by the retrieval delay ($\tau_\text{ret}$) in Fig.~\ref{fig4}. Solid curves are fits to either exponential or Gaussian decay (see Methods section \ref{lifetime_sec}). Dotted horizontal line denotes the $1/e$ point for total efficiency.
 }
	\label{fig5}
\end{figure}

In Fig.~\ref{fig5}, we provide the data and model fits used to extract the memory lifetimes shown in Fig.~\ref{fig2}b. Each lifetime measurement is performed by increasing the argon buffer gas pressure inside the heat pipe to a target setpoint and measuring the total (end-to-end) memory efficiency as a function of varying retrieval delay, as shown in Fig.~\ref{fig4}. We fit each set of data to either a single exponential decay or a single Gaussian decay, depending on whether the argon pressure is above or below an arbitrary cutoff of 25 mbar. Above this threshold, the experimental data are well described by exponential decay, and below this threshold, the experimental data are better fit to a Gaussian decay function characteristic of Doppler dephasing. The error bars in Fig.~\ref{fig5} represent systematic uncertainty due to control field power and beam pointing fluctuations, which can be quite large in our experiment, and the propagated error in Fig.~\ref{fig2}b represents the fitting uncertainty in the presence of these systematics.

\subsection{Maxwell-Bloch equations.}

We model our experiment using the well-known Maxwell-Bloch equations \cite{scully1997quantum,gorshkov2007universal,gorshkov2007photon_2,nunn2008quantum,shinbrough2021optimization}:

\begin{align}
    \label{Aeq}\partial_z A(z,\tau) &= -\sqrt{d} P(z,\tau)\\
    \label{Peq}\partial_\tau P(z,\tau) &= -\bar{\gamma} P(z,\tau) + \sqrt{d} A(z,\tau) - i\frac{\Omega(\tau)}{2} B(z,\tau)\\
    \label{Beq}\partial_\tau B(z,\tau) &= -\gamma_B B(z,\tau) -i\frac{\Omega^*(\tau)}{2} P(z,\tau),
\end{align}

\noindent where $z$ represents the one-dimensional spatial coordinate of the atomic ensemble normalized to the ensemble length [i.e., $z=0$ ($z=1$) represents the beginning (end) of the ensemble]; $\tau = t-z/c$ represents time measured in the comoving frame of the signal photon ($t$ represents time in the lab frame) normalized to the excited-state coherence decay rate $\gamma = \Gamma/2$ ($\Gamma$ is the total excited-state population decay rate, or the linewidth of the $\ket{g}\leftrightarrow\ket{e}$ transition); $A(z,\tau)$ is the spatially and temporally dependent signal photonic field; $P(z,\tau)$ and $B(z,\tau)$, referred to as the atomic polarization and spin wave fields, respectively, are macroscopic field operators representing the atomic coherences $\ket{g}\leftrightarrow\ket{e}$ and $\ket{g}\leftrightarrow\ket{s}$, which are delocalized across the length of the medium and are shown in Fig.~\ref{fig1}b as orange and blue shaded regions, respectively; $d$ is the resonant optical depth of the memory; $\bar{\gamma} = (\gamma-i\Delta)/\gamma$ is the normalized complex detuning, where the detuning $\Delta$ is shown schematically in Fig.~\ref{fig1}b; and $\Omega(\tau)$ is the control field Rabi frequency coupling the $\ket{e}$ and $\ket{s}$ states. All atomic population is assumed to start in the ground state, and the metastable storage state is assumed to have a coherence decay rate $\gamma_B$ that is much smaller than the excited state decay rate ($\gamma_B\ll1$, in normalized units).

The set of equations \eqref{Aeq}-\eqref{Beq} define a map between the input photonic field $A(z=0,\tau)$ and either the output spin-wave field $B(z,\tau\rightarrow\infty)$, in the case of the quantum storage operation, or the output photonic field $A(z=1,\tau)$, in the case of storage and retrieval operations. In quantum storage, the storage efficiency is defined as $\eta = \int_{0}^{1}dz \abs{B(z,\infty)}^2 \big/ \int_{-\infty}^{\infty}d\tau \abs{A(0,\tau)}^2$, which is well-approximated by $\eta \approx \int_{-\infty}^{\infty}d\tau \abs{A(1,\tau)}^2 \big/ \int_{-\infty}^{\infty}d\tau \abs{A(0,\tau)}^2$ when all of the signal field population entering the atomic system is transferred into the storage state (i.e., no spontaneous emission loss). This condition is met, for example, in the absorb-then-transfer protocol when there is unit efficiency transfer between $P$ and $B$ fields (ensured by a $\pi$-pulse-area control field) and no excited-state decay during the storage operation (ensured when the storage-control-field delay $\Deltau$ is much shorter than the decay time, $\Gamma$, which is trivially the case in the broadband regime when the pulse duration $\tau_\text{FWHM}\ll 1/\Gamma$ and $\Deltau\sim\tau_\text{FWHM}$). This condition is also met in the idealized EIT regime with complete adiabatic elimination of the atomic polarization field, and in the off-resonant regime when the detuning is sufficiently large that no linear absorption takes place in the absence of the control field, and the presence of the control field maps population only to the storage state. This approximation to $\eta$ is useful as it allows for measurement of storage efficiency through photon counting alone. After storage and retrieval, the total efficiency $\eta_\text{tot} = \int_{-\infty}^{\infty}d\tau \abs{A(1,\tau>\tau_\text{ret})}^2 \big/ \int_{-\infty}^{\infty}d\tau \abs{A(0,\tau)}^2$, where $\tau_\text{ret}$ is the retrieval delay or storage time, can also be measured via photon counting by subtracting the photon counts corresponding to transmission during the storage operation (`leaked' photons) from the total output photon counts.

Eq. \eqref{Aeq}-\eqref{Beq} can also be used as a numerical fitting function.
The fits performed in Fig.~\ref{fig2}a and \ref{fig2}c assume Fourier-limited signal and control fields with experimentally measured bandwidth, control field power, signal--control-field delay, detuning, and waist radii; literature values for the control field transition dipole matrix element, and the signal and control field center frequencies; and use the optical depth and excited-state coherence decay rate as fit parameters.

\subsection{Near-Off-Resonant Memory (NORM) operation.}\label{NORM_sec}

\begin{figure}[t]
	\centering
	\includegraphics[width=0.85\columnwidth]{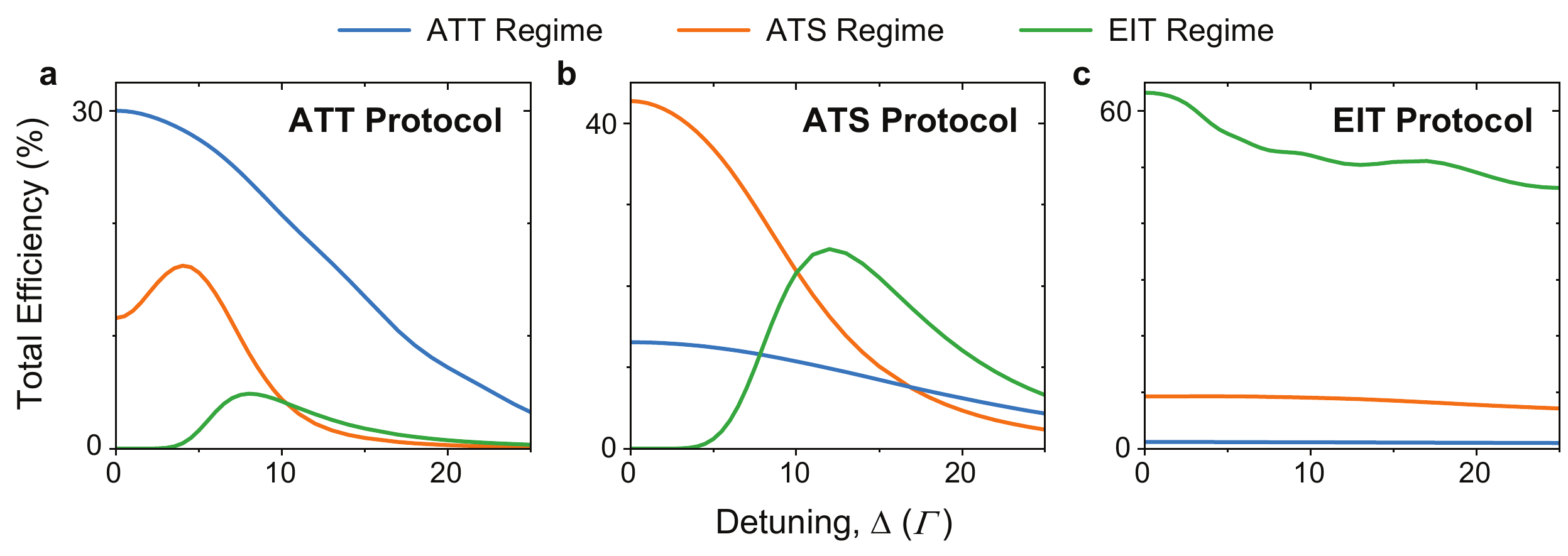}
	\caption{\textbf{Simulations of near-off-resonant-memory operation.} \textbf{a}-\textbf{c}, 
    Total memory efficiencies plotted as a function of detuning for the absorb-then-transfer (ATT) (\textbf{a}), Autler-Townes splitting (ATS) (\textbf{b}), and electromagnetically induced transparency (EIT) (\textbf{c}) protocols. Solid curves show the memory efficiencies for each protocol when applied in the ATT (blue), ATS (orange), and EIT (green) memory parameter regimes (see Methods section \ref{NORM_sec}). Maximal efficiency occurs at non-zero detuning when the memory adiabaticity (dependent on memory regime) is larger than the effective adiabaticity for a particular control field (dependent on memory protocol).
 }
	\label{fig6}
\end{figure}

It may be unexpected \textit{a priori} that an ensemble-based memory with a given optical depth, linewidth, and control field parameters possesses can possess a finite, non-zero optimal detuning. Especially in the ultrabroadband regime considered in this work, where we employ the absorb-then-transfer protocol that is limited in memory efficiency by linear absorption, one would na\"{i}vely expect the maximal memory efficiency always occurs on resonance, where linear absorption is maximized. Here we aim to provide some physical intuition as to why this is not the case.

Our intuition relies on understanding the control fields used in ensemble-based memory to possess their own adiabaticity, distinct from the memory adiabaticity. As in the rest of the literature, we define the free-space memory adiabaticity as $\chi = d\tau_\text{FWHM}\gamma$ 
\cite{gorshkov2007universal,gorshkov2007photon_2,nunn2008quantum}, and gives an indication of how slowly varying the signal field is relative to the response of the medium defined by $d$ and $\gamma$. As in Refs.~\cite{shinbrough2021optimization,shinbrough2023variance}, we define the memory parameters $\mathcal{M} \equiv (d,\tFWHM\gamma)$, and for simplicity we limit our analysis to the case of Gaussian control fields, where the optimal Gaussian control field for a given set of memory parameters is uniquely defined by $\mathcal{G}(\mathcal{M})$. For a given control field, we can invert this function and find the corresponding optimal memory parameters $\mathcal{M}'(\mathcal{G}) = (d',\tFWHM\gamma')$ and effective adiabaticity for a particular control field $\chi'$
. When $\chi'<\chi$ (the control-field adiabaticity is less than the memory adiabaticity) we observe NORM operation --- a maximal total memory efficiency at finite, non-zero detuning. 
In this case, the memory's linear absorption is stronger than what is optimal for the control field being used, and it is beneficial to increase the detuning slightly to effectively reduce the linear absorption of the memory.

Another way to explain this behavior is in terms of reabsorption loss. In optically thick atomic ensembles, some fraction of the retrieved signal field must propagate through the atomic ensemble before reaching the free-space output port of the memory. This propagation introduces additional loss to the memory operation, as the retrieved signal field has some probability of being re-absorbed by the ensemble. This is a well-known source of memory inefficiency \cite{afzelius2009multimode}, and is strongest on resonance where the signal field overlaps with the absorption spectrum of the ensemble. It is therefore often beneficial to detune from resonance to avoid reabsorption loss, but increasing the detuning without increasing the control field strength can also decrease the light-matter coupling and therefore the memory efficiency. A tradeoff between these two effects leads to a maximal memory efficiency at non-zero two-photon detuning. We note, however, that this explanation is incomplete as it does not account for a control field sufficiently strong that it opens a transparency window at the signal frequency and eliminates reabsorption loss. The description above in terms of memory and control-field adiabaticities is more general and complete.

In Fig.~\ref{fig6}, we numerically simulate several combinations of memory parameters and control field parameters that elucidate NORM operation and verify our intuition. We choose three sets of memory parameters that correspond to the three resonant memory regimes: $\mathcal{M} = (5,0.1)$, $(7.5,0.4)$, and $(50,1.5)$ for the ATT, ATS, and EIT regimes, respectively. For each set of memory parameters, we find the three sets of unique, optimal Gaussian control field parameters $\mathcal{G}(\mathcal{M}) = \left(\theta,\Deltau,\tctrl\right)$, where $\theta$ is the control field pulse area (units of $\pi$), $\Deltau$ is the control field delay relative to the signal field (units of $\tFWHM$), and $\tctrl$ is the control field duration (intensity FWHM, units of $\tFWHM$). As the control field parameters define the memory protocol, each $\mathcal{G}(\mathcal{M})$ constitutes a different memory protocol; we use $\mathcal{G}(\mathcal{M}) = (1.0789,0.76176,0.52137)$, $(2.63177,-0.23817,1.23829)$, and $(10.05845,-0.54359,1.33658)$ for the ATT, ATS, and EIT protocols, respectively. Fig~\ref{fig6}a shows numerical applications of the ATT protocol in the ATT (blue), ATS (orange), and EIT (green) regimes, and Fig~\ref{fig6}b and c show the same for the ATS and EIT protocols, respectively. Clear near-off-resonant-memory operation occurs in Fig~\ref{fig6}a and b when applying a less adiabatic memory protocol in a more adiabatic memory regime. In some cases, e.g., when using the ATT and ATS protocols in the EIT regime, the memory efficiency is near zero on resonance and only increases to appreciable values for near-resonant detunings. As one would expect, each memory protocol is most efficient in its respective regime.

Our experiment and the data shown in Fig.~\ref{fig2}c are most similar to the ATS regime in Fig.~\ref{fig6}a, where we employ the ATT protocol (due to the available control field power and pulse duration) in a regime more suitable to the ATS (or a mixed ATT-ATS) protocol.

\subsection{Amplitude and phase reconstruction via spectral interferometry.}

\begin{figure}[t]
	\centering
	\includegraphics[width=1\columnwidth]{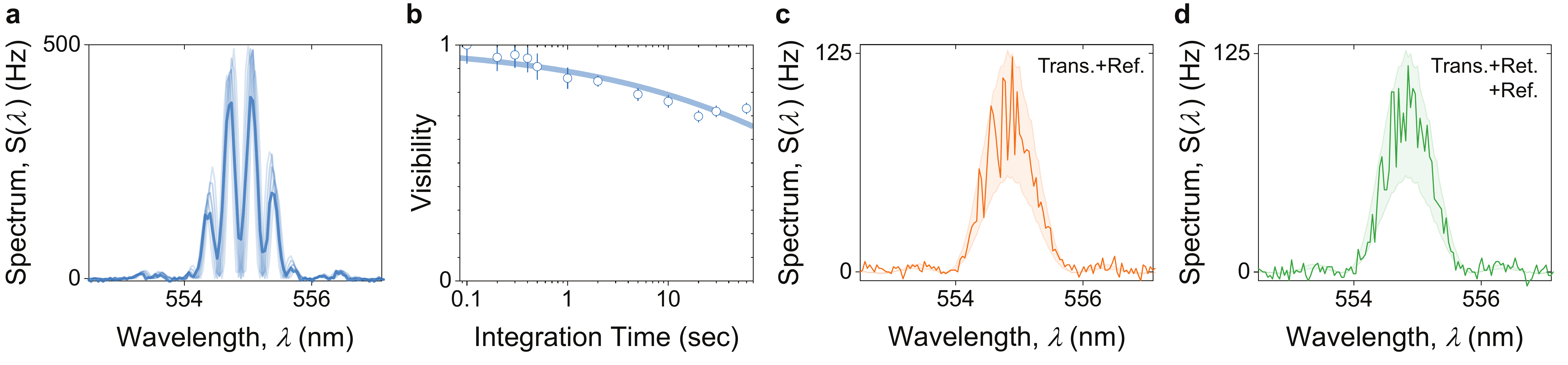}
	\caption{\textbf{Spectral Interferometry.} \textbf{a}, Non-unit visibility measured spectral interference between the two interferometer paths with the memory off (solid blue curve) occurs at finite integration time due to averaging of many short-duration interference spectra with varying phase (shaded blue curves). \textbf{b}, Measured spectral interference visiblity (markers) as a function of spectrometer integration time. Error bars represent statistical error from photon counting statistics propagated through a sinusoidal fit. Solid curve is a fit to the data based on a time-varying Gaussian phase distribution (see Methods). \textbf{c}-\textbf{d}, Spectral interference between the transmitted and reference (\textbf{c}), and transmitted, retrieved, and reference (\textbf{d}) pulses with the memory on. Shaded regions designate the expected bounds for non-unit interference visiblity according to our model (see Methods).
 }
	\label{fig7}
\end{figure}

In order to reconstruct the retrieved signal field amplitude and phase from spectral interference measurements, we first consider the general case of two ultrafast pulses with electric fields $A_1(\tau)$ [$A_1(\omega)$] and $A_2(\tau)$  [$A_2(\omega)$] in the temporal (spectral) domain. We take $A_1(\tau)$ to be a reference pulse with known amplitude and phase and $A_2(\tau)$ to be a modified pulse similar in spectral bandwidth to $A_1(\tau)$ but with differing amplitude $\abs{A_2(\tau)}$ and temporal phase $\phi_2(\tau)$, which we aim to measure. If we combine $A_1(\tau)$ and $A_2(\tau)$ on a beamsplitter with time delay $\Delta \tau$, the resulting interference spectrum in one output port of the beamsplitter can be recorded on a spectrometer and used to reconstruct $A_2(\tau)$. The details of this process are as follows.

The Fourier decomposition of $A_j(\tau)$ (for $j=1,2$) is given by $A_j(\tau) = \frac{1}{\sqrt{2\pi}} \int_{-\infty}^{\infty} d\omega \, \abs{A_j(\omega)}e^{i[\omega \tau + \phi_j(\omega)]}$, 
where $A_j(\omega) = \abs{A_j(\omega)}e^{i \phi_j(\omega)}$. The Fourier decomposition of the signal impinging on the spectrometer is therefore: 
$\frac{1}{2\sqrt{\pi}} \int_{-\infty}^{\infty} d\omega \, \big[i \abs{A_1(\omega)}e^{i\phi_1(\omega)} + \abs{A_2(\omega)}e^{i[\omega \Delta \tau + \phi_2(\omega)]} \big]e^{i\omega \tau}$,
where we have assumed (without loss of generality) that $A_1(\tau)$ is reflected at the beamsplitter and $A_2(\tau)$ is transmitted. The spectrometer detects the spectral intensity of the signal impinging on it:
\begin{equation}
	\label{S}S(\omega) 
= \abs{A_1(\omega)}^2 + \abs{A_2(\omega)}^2 + 2\abs{A_1(\omega)}\abs{A_2(\omega)}\sin{[\omega \Delta \tau + \phi_\text{dif}(\omega)]},
\end{equation}

\noindent where $\phi_\text{dif}(\omega) = \phi_2(\omega)-\phi_1(\omega)$. Both $\abs{A_1(\omega)}^2$ and $\abs{A_2(\omega)}^2$ (and therefore $\abs{A_1(\omega)}$ and $\abs{A_2(\omega)}$ by applying the square root) can be measured trivially by blocking the opposite input port of the beamsplitter and recording the single-path spectrum. If $\Delta \tau$ is known, and is smaller than the inverse of the spectrometer resolution, everything in Eq.~\eqref{S} is known except for $\phi_\text{dif}(\omega)$. We can therefore measure the interferogram $S(\omega)$, fit to Eq.~\eqref{S}, extract $\phi_\text{dif}(\omega)$ [and therefore $\phi_2(\omega)$ for known $\phi_1(\omega)$], and reconstruct $A_2(\omega)$ and  $A_2(\tau)$ using the relations above.
	
In experiment, we follow this procedure by splitting the signal field as generated via sum-frequency generation (see Sec.~\ref{exp_setup_sec}) into two paths; one path is sent to the barium heat pipe quantum memory for storage and retrieval, and the other path propagates in free space and acts as a reference. The two paths are recombined on a beamsplitter before being sent to a single-photon-level spectrometer with $\sim$0.04 nm resolution. With the memory off (heat pipe at 20 $^\circ$C, where the number density of gas phase barium atoms is negligible), we observe spectral interference between the two paths as shown in Fig.~\ref{fig7}a. At spectrometer integration times less than 0.1 sec we achieve near unit visibility spectral interference, but the statistical error do to photon counting statistics is significant and makes spectral reconstruction of the signal field with the memory on impossible. We must therefore consider longer spectrometer integration times, where the interference visibility is significantly below unity. Measured interference visibilities as a function of spectrometer integration time are shown in Fig.~\ref{fig7}b, along with a model fit derived from first principles: We define the time-averaged visibility $\overline{V} = (\overline{I_\text{max}}-\overline{I_\text{min}})/(\overline{I_\text{max}}+\overline{I_\text{min}})$ in terms of the time-averaged maximum and minimum intensities, $\overline{I_\text{max}}$ and $\overline{I_\text{min}}$, respectively. We assume that one can convert the spectral interference in, e.g., Fig.~\ref{fig7}a into a purely sinusoidal interference spectrum and an envelope function (which, in experiment, we measure by blocking one interferometer path at a time and summing the resulting single-path spectra). In this case, the time-averaged maximum and minimum sinusoid intensities are $\overline{I_\text{max}} = \int d\phi \, P(\phi,t) \sin^2(\phi + \pi/2)$ and $\overline{I_\text{min}} = \int d\phi \, P(\phi,t) \sin^2(\phi)$, for a phase distribution $P(\phi,t)$ at integration time $t$. For a Gaussian phase distribution, $P(\phi,t) \sim e^{-\phi^2/[2\sigma(t)^2]}$, the time-dependence is contained solely in the parameter $\sigma(t)$, and leads to an time-averaged visibility $\overline{V} = e^{-2\sigma(t)^2}$. The exact time dependence of $\sigma(t)$ depends on the specific experimental apparatus. We achieve reasonable agreement with experimental data using the general form $\sigma(t) = f_1 t^{f_2}$ for fit parameters $f_1$ and $f_2$, which evaluate to $f_1 = 0.06$ and $f_2 = 0.3$ in our experiment, as shown in Fig.~\ref{fig7}b. Using this model for non-unit visibility interference, along with the measured spectral intensities of the transmitted, retrieved, and reference fields, we estimate the expected upper and lower bounds for non-unit visibility spectral interference shown in Fig.~\ref{fig7}c and d.

Fig.~\ref{fig7}c shows the measured interference spectrum after combining the transmitted and reference pulses on a beamsplitter after the memory. We isolate the transmitted pulse from the retrieved pulse experimentally by blocking the retrieval control field. The interference spectrum shows significantly asymmetrical visibility, as expected for a transmitted pulse with linear temporal phase relative to the reference pulse. When unblocking the retrieval control field, we observe the interference spectrum in Fig.~\ref{fig7}d, which again has an asymmetrical interference visibility, but with a significantly more complicated interference pattern. We use the method described above to reconstruct the amplitude and phase (relative to the reference pulse) of the transmitted and retrieved signal field pulses, leading to the results shown in Fig.~\ref{fig3}c.

\subsection{Signal-field frequency dependence.}

\begin{figure}[t]
	\centering
	\includegraphics[width=0.85\columnwidth]{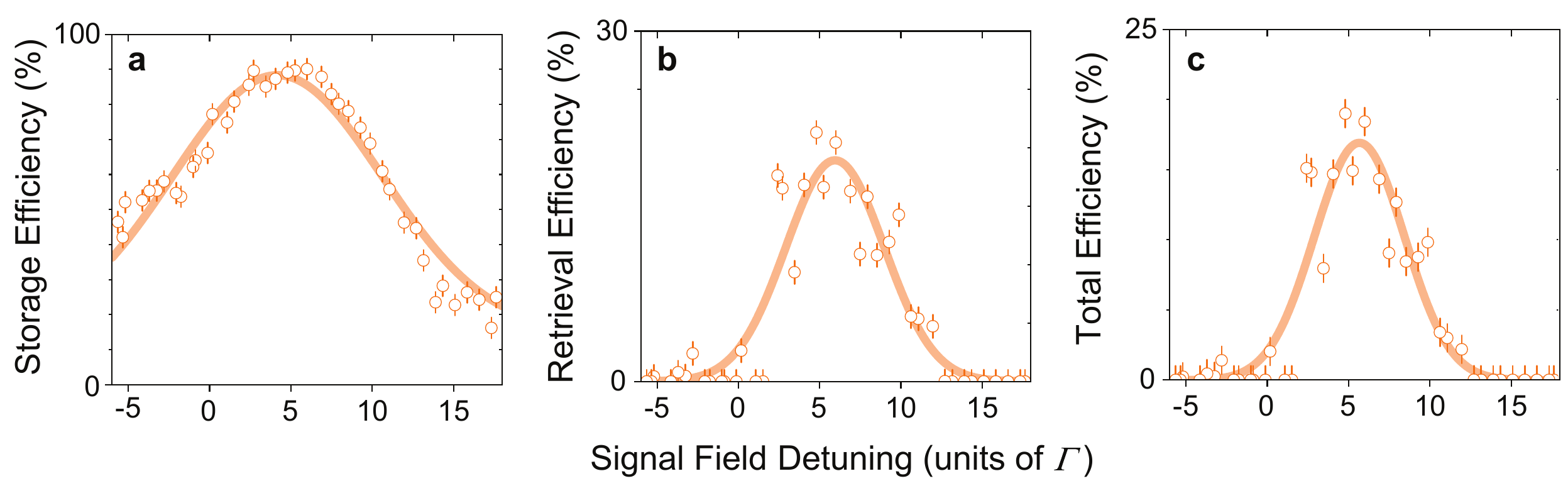}
	\caption{\textbf{Signal field frequency dependence.} \textbf{a}-\textbf{c}, Measured storage (\textbf{a}), retrieval (\textbf{b}), and total (\textbf{c}) efficiencies as a function of signal field center frequency, written as the detuning from resonance. Solid lines are fits representative of the light-matter coupling due to the Gaussian control field centered near a detuning of $5\Gamma$. Error bars represent systematic error due to control field power variation.
 }
	\label{fig8}
\end{figure}

As a further characterization step, we measure the signal-field frequency dependence of our memory. We keep the center frequency of our control field fixed at around $5\Gamma$ off resonance and vary the center frequency of our signal field, scanning over the near-resonant coupling spectrum due to our Gaussian control field. In Fig.~\ref{fig8} we report the measured storage, retrieval, and total memory efficiencies along with fits to independent Gaussian functions. The storage efficiency spectrum is approximately a factor of $\sqrt{2}$ wider than the retrieval and total efficiency spectra as the storage operation requires the action of a single control field pulse, whereas the retrieval (and therefore total) operation requires the action of two control field pulses.


\bibliography{sn-bibliography}

\end{document}